\documentclass[reprint,
preprintnumbers,
amsmath,amssymb,
]{revtex4-1}

\makeatletter
\let\frontmatter@footnote@produce\frontmatter@footnote@produce@endnote
\makeatother

\usepackage{graphicx}% Include figure files
\usepackage{dcolumn}% Align table columns on decimal point
\usepackage{bm}% bold math
\usepackage{amssymb,physics,amsmath}
\usepackage{braket}
\usepackage[colorlinks,allcolors=blue]{hyperref}% add hypertext capabilities
\def\bk{\bm{k}}
\def\bkp{\bm{k}'}
\def\bR{\bm{R}}

\def\bRp{\bm{R}'}
\def\bkc{\bm{k}_{\rm c}}
\def\bkpc{\bm{k}'_{\rm c}}
\def\bkf{\bm{k}_{\rm f}}
\def\bkpf{\bm{k}'_{\rm f}}

\begin{document}

\title{\textit{Ab initio} electron-defect interactions using Wannier functions}

\author{I-Te Lu}
\author{Jinsoo Park}
\author{Jin-Jian Zhou}%
\author{Marco Bernardi}
\email{bmarco@caltech.edu}
\affiliation{Department of Applied Physics and Materials Science, California Institute of Technology, Pasadena, California 91125}

%\date{\today}

%%%%%%%%%%%%%%%%%%%%%%%%%%%%%%
% ABSTRACT
%%%%%%%%%%%%%%%%%%%%%%%%%%%%%%
\begin{abstract}
\noindent
% 
% Opening and challenge 
Computing electron-defect ($e$-d) interactions from first principles has remained impractical due to computational cost. 
%
% Action
Here we develop an interpolation scheme based on maximally localized Wannier functions (WFs) to efficiently compute $e$-d interaction matrix elements. 
%
% Resolution
The interpolated matrix elements can accurately reproduce those computed directly without interpolation, and the approach can significantly speed up calculations of $e$-d relaxation times and defect-limited charge transport.
We show example calculations of vacancy defects in silicon and copper, for which we compute the $e$-d relaxation times on fine uniform and random Brillouin zone grids 
(and for copper, directly on the Fermi surface) as well as the defect-limited resistivity at low temperature. 
% 
% Applications
Our interpolation approach opens doors for atomistic calculations of charge carrier dynamics in the presence of defects.

\end{abstract}

\maketitle

%%%%%%%%%%%%%%%%%%%%%%%%%%%%%%
% INTRODUCTION
%%%%%%%%%%%%%%%%%%%%%%%%%%%%%%
\noindent
\textbf{INTRODUCTION}\\
% 
% THE IMPORTANCE OF ELECTRON-DEFECT INTERACTIONS AND THE MAIN ISSUE
%
The interactions between electrons and defects control charge and spin transport at low temperature, and give rise to a range of quantum transport phenomena~\cite{economou1983green,datta1997electronic,bruus2004many,akkermans2007mesoscopic,mahan2013many}.
%~\cite{Shiranzaei2017,Zhou2016,Kim2014,Jirauschek2014,Marconcini2012,Persson2010}
%~\cite{Strozecka2011,Lu2011}. 
%
Understanding such electron-defect ($e$-d) interactions from first principles can provide microscopic insight into carrier dynamics in materials in the presence of point and extended defects, and accelerate materials discovery. 
Currently, $e$-d interactions can be computed \textit{ab initio} mainly through the all-electron Korringa-Kohn-Rostoker Green's function method~\cite{ebert2011calculating}, which is based on multiple-scattering theory.
Several properties have been computed with this approach, including the magnetoresistance due to impurity scattering~\cite{zahn1995ab}, the residual resistivity of metals and alloys~\cite{mertig1999transport}, 
and spin relaxation~\cite{fedorov2008first,gradhand2010fully,fedorov2013impact,long2013spin} and the spin Hall effect~\cite{gradhand2010spin,gradhand2010extrinsic,honemann2019spin}. %, among others
In contrast, \textit{ab initio} $e$-d calculations using pseudopotentials or projector augmented waves~\cite{Restrepo2009,Lordi2010,Lu2019} have seen slower progress, 
mainly due to the high computational cost of obtaining the $e$-d interaction matrix elements needed for perturbative calculations~\cite{bernardi2016first}. 
%

%
% COMPARSION PW AND AO => INTERPOLATION SCHEME
%
% our method
We recently developed an \textit{ab initio} method~\cite{Lu2019} to compute efficiently the $e$-d interactions and the associated matrix elements. 
Our approach uses only the wave functions of the primitive cell, thus significantly reducing computational cost compared to $e$-d calculations that use supercell wave functions~\cite{Restrepo2009,Lordi2010}. 
Since our method uses a plane wave (PW) basis set and pseudopotentials, it is compatible with widely used density functional theory (DFT) codes. 
%
% AO method
A different method developed by Kaasbjerg \textit{et al.}~\cite{Kaasbjerg2017} uses an atomic orbital (AO) basis to compute the $e$-d matrix elements; the advantage of this approach is 
that one can compute the $e$-d matrix elements using only a small set of AOs, although one is limited by the quality and completeness of the AO basis set~\cite{Ulian2013}.\\ %are an intrinsic limit of the method 
\indent
To benefit from both the completeness and accuracy of the PW basis and the versatility of a small localized basis set, approaches combining PWs and AOs~\cite{Agapito2018} or Wannier functions (WFs)~\cite{Giustino2007} have been developed for electron-phonon ($e$-ph) interactions. They have enabled efficient interpolation of the $e$-ph matrix elements and have been instrumental to advancing carrier dynamics calculations~\cite{ZhouBernardi2016, Zhou2018,Lee2018}.
%
% gap sentence
To date, such an interpolation scheme does not exist for $e$-d interactions to our knowledge, so performing demanding Brillouin zone (BZ) integrals needed to compute $e$-d relaxation times (RTs) and defect-limited charge transport remains an open problem. 
Interpolating the interaction matrix elements to uniform, random, or importance-sampling fine BZ grids is key to systematically converging the RTs and transport properties~\cite{ZhouBernardi2016,park2019elliott}, and it has been an important development in first-principles calculations of $e$-ph interactions and phonon-limited charge transport.

%
% PURPOSE AND CONTRIBUTION OF THIS WORK
%
In this work, we develop a method for interpolating the $e$-d interaction matrix elements using WFs.
Through a generalized double-Fourier transform, our approach can efficiently transform the matrix elements from a Bloch representation on a coarse BZ grid to a localized WF representation 
and ultimately to a Bloch representation on an arbitrary fine BZ grid. 
We show the rapid spatial decay of the $e$-d interactions in the WF basis, which is crucial to the accuracy and efficiency of the method.
Using our approach, we investigate $e$-d interactions due to charge-neutral vacancies in silicon and copper. 
%calculations of 
In both cases, we can accurately interpolate the $e$-d matrix elements and converge the $e$-d scattering rates and defect-limited carrier mobility or resistivity. 
In copper, we map the $e$-d RTs directly on the Fermi surface, and show their peculiar dependence on electronic state.
We demonstrate computations of $e$-d matrix elements on random and uniform BZ grids as dense as $600\times600\times600$ points, whose computational cost would be prohibitive for direct computation.
We expect that our efficient $e$-d computation and interpolation approaches will enable a wide range of studies of $e$-d interactions in materials ranging from metals to semiconductors and insulators and for applications such as electronics, nanodevices, spintronics and quantum technologies.\\

\noindent
\textbf{RESULTS}\vspace{2pt}\\
\textbf{Theory}\vspace{2pt}\\
%
% INTORUDCE E-D MATRIX ELEMENTS   
%
% Bloch representatino
The perturbation potential $\Delta V_{\rm e-d}$ introduced by a point defect in a crystal couples different Bloch eigenstates of the unpertured (defect-free) crystal. 
The matrix elements associated with this $e$-d interaction are defined as
%
%EQ: e-d in bloch representation
\begin{equation}\label{eq:ed-vertex-bloch}
M_{mn}(\bkp,\bk)=\braket{m\bkp\,|\,\Delta V_{\rm e-d}\,|\,n\bk},
\end{equation}
where $\ket{n\bk}$ is the Bloch state with band index $n$ and crystal momentum $\bk$.
To handle these $e$-d interactions, one needs to store and manipulate a matrix $M_{mn}$ of size $N_{b}^{2}$ ($N_{b}$ is the number of bands) for each pair of crystal momenta $\bkp$ and $\bk$ in the BZ.
Within DFT, the perturbation potential $\Delta V_{\rm e-d}$ can be computed as the difference between the Kohn-Sham potential of a defect-containing supercell and that of a pristine supercell with no defect~\cite{Lu2019}.\\
\indent
We compute the $e$-d matrix elements in Eq.~(\ref{eq:ed-vertex-bloch}) using the method we developed in Ref.~\cite{Lu2019}, 
which uses only the Bloch wave functions of the primitive cell and does not require computing or manipulating the wave functions of the supercell, thus significantly reducing computational cost.
%
% introduce Bloch states 
The Bloch states can be expressed in terms of maximally localized WFs using
%
%EQ: Bloch state <- Wannier functions
\begin{equation}\label{eq:w2bfunct}
\ket{n\bk}=\sum_{j\bR}e^{i\bk\cdot\bR}U^{\dagger}_{jn,\bk}\ket{j\bR},
\end{equation}
where $\ket{j\bR}$ is the WF with index $j$ centered at the Bravais lattice vector $\bR$. 
%
% introduce Wannier functions 
The unitary matrices $U$ in Eq.~(\ref{eq:w2bfunct}) maximize the spatial localization of the WFs~\cite{Marzari1997}
%
%EQ: Wannier functions <- Bloch states
\begin{equation}\label{eq:b2wfunct}
\ket{j\bR}=\frac{1}{N_{\bk}}\sum_{n\bk}e^{-i\bk\cdot\bR}U_{nj,\bk}\ket{n\bk},
\end{equation}
where $N_{\bk}$ is the number of $\bk$-points in the BZ. 
%
% Wanniner representation
The $e$-d matrix element $M_{ij}(\bRp,\bR)$ between two WFs centered at the lattice vectors $\bRp$ and $\bR$ in the Wannier representation is defined as  
%
%EQ: e-d in Wannier representation
\begin{equation}\label{eq:ed-vertex-wannier}
M_{ij}(\bRp,\bR)= \braket{i\bRp\,|\,\Delta V_{\rm e-d}\,|\,j\bR}.
\end{equation}
If the center of the perturbation potential lies in the unit cell at the origin, the absolute value of the $e$-d matrix elements, $|M_{ij}(\bRp,\bR)|$, decays rapidly (within a few lattice constants) for increasing values of the lattice vectors $\bRp$ and $\bR$ due to the short-ranged nature of the perturbation potential from the defect, which is assumed to be charge-neutral in this work.
As a result, only a small number of lattice vectors $\bR$, which we arrange in a Wigner-Seitz (WS) supercell centered at the origin, is needed to compute the $e$-d matrix elements in the Wannier representation.

%
% FOURIER-WANNIER INTERPOLATION FORMULA
%
Using Eqs.~(\ref{eq:ed-vertex-bloch})-(\ref{eq:ed-vertex-wannier}), the $e$-d matrix elements in the Wannier representation can be written as a generalized double Fourier transform of the matrix elements in the Bloch representation, 
which are first computed on a coarse BZ grid with points $\bkc$:
\begin{equation}\label{eq:ed-b2w}
M(\bRp,\bR)=\left(\frac{1}{N_{\bkc}}\right)^{2}\sum_{\bkpc\bkc}e^{i(\bkpc\cdot\bRp-\bkc\cdot\bR)}U^{\dagger}_{\bkpc}M(\bkpc,\bkc)U_{\bkc}.
\end{equation}
Here and below, we omit all band indices for clarity.
Through the inverse transform, we can interpolate the $e$-d matrix elements to any desired pair of fine BZ grid points $\bkpf$ and $\bkf$, using
\begin{equation}\label{eq:ed-w2b}
M(\bkpf,\bkf)=\sum_{\bRp\bR}e^{-i(\bkpf\cdot\bRp-\bkf\cdot\bR)}U_{\bkpf}M(\bRp,\bR)U^{\dagger}_{\bkf}.
\end{equation}
While $U_{\bkc}$ in Eq.~(\ref{eq:ed-b2w}) is the coarse-grid unitary matrix used to construct the WFs [see Eq.~(\ref{eq:b2wfunct})], the unitary matrix on the fine grid, $U_{\bkf}$ in Eq.~(\ref{eq:ed-w2b}), is obtained by diagonalizing the fine-grid Hamiltonian,
\begin{equation}\label{eq:hr2hk}
H(\bkf)=\sum_{\bR}e^{i\bkf\cdot\bR}H(\bR),
\end{equation}
where $H(\bR)$ is the electronic Hamiltonian in the Wannier basis.
%
% Similarity with the e-ph formula
These equations are analogous to those used for interpolating the $e$-ph matrix elements~\cite{Giustino2007,Agapito2018}, except that here the lattice vectors $\bRp$ and $\bR$ are both associated with electronic states.
The lattice vectors $\bRp$ and $\bR$ in the WS supercell are determined$-$through the periodic boundary conditions$-$by the $\bkpc$ and $\bkc$ coarse grids, respectively. 
In practice, we choose a uniform coarse BZ grid, and the size of the WS supercell is equal to the size of this coarse grid.\\
\indent 
Similar to the $e$-ph case~\cite{Agapito2018}, the rapid spatial decay of the $e$-d matrix elements in the WF basis is crucial to reducing the computational cost 
since it puts an upper bound to the number of lattice sites $\bRp$ and $\bR$ at which $M(\bRp,\bR)$ needs to be computed.
In particular, while computing $M(\bkpf,\bkf)$ at small $\bkpf$ and $\bkf$ vectors would in principle require summing the Fourier transform in Eq.~(\ref{eq:ed-w2b}) 
up to correspondingly large lattice vectors of length $|\bRp|=2\pi/|\bkpf|$ and $|\bR|=2\pi/|\bkf|$, in practice this is not needed due to the rapid spatial decay.
The choice of a WS supercell and its relation to the DFT supercell are discussed in detail in the Supplementary Information. \\

%
%%%%%%%%%%%%%%%%%%%%%%%%
%%% Figure 1: Workflow 
%%%%%%%%%%%%%%%%%%%%%%%%
\begin{figure}[t!]
\centering %0.85
\includegraphics[width=\linewidth]{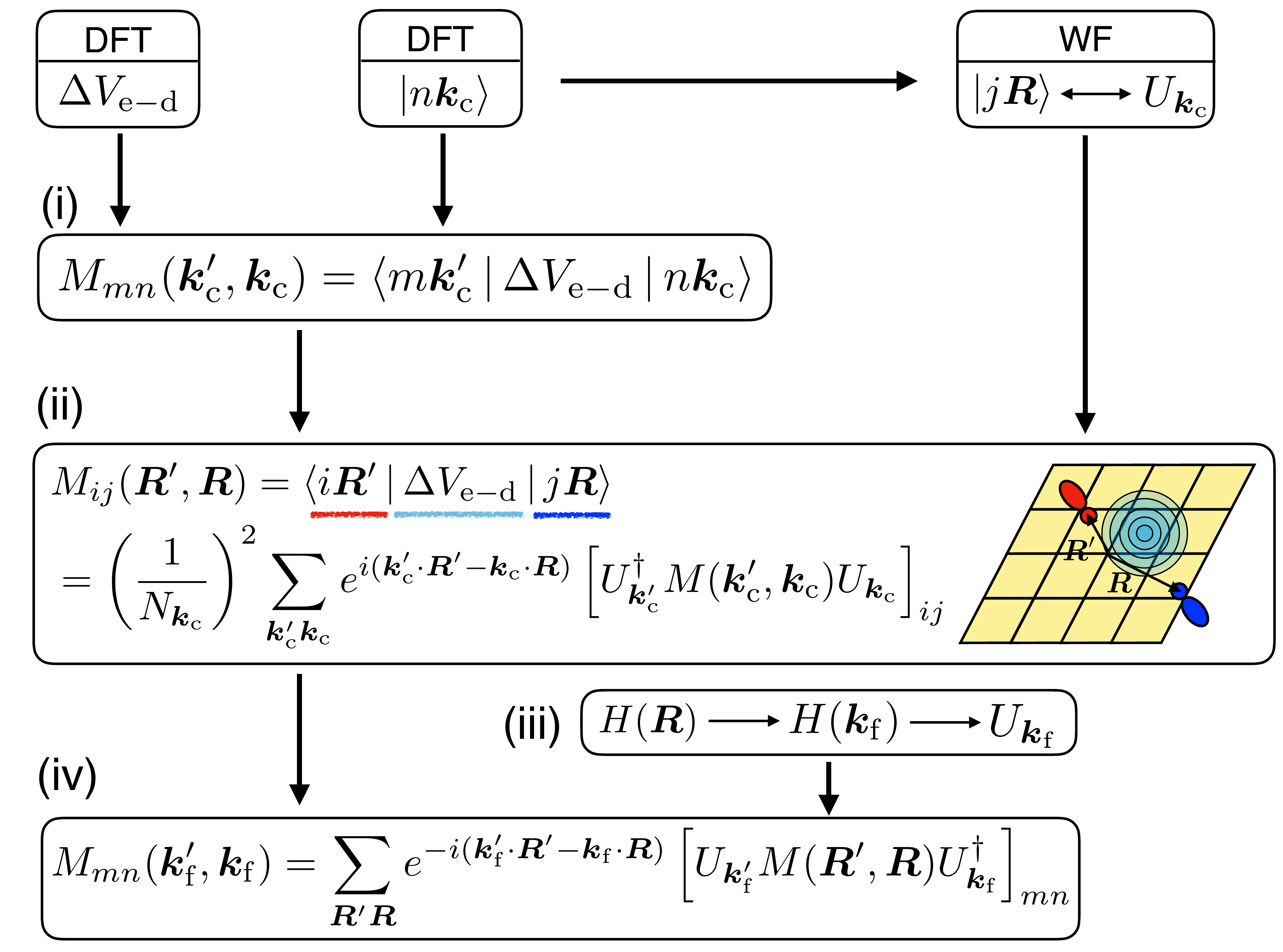}
\caption{
Workflow for interpolating the $e$-d matrix elements using WFs. The steps are numbered as in the text.
}\label{fig:workflow}
\end{figure}

\noindent
\textbf{Interpolation workflow and validation}\vspace{2pt}\\
%
% WORKFLOW FOR THE WANNIER INTERPOLATION OF E-D MATRIX ELEMENTS
%
The workflow for interpolating the $e$-d matrix elements to a fine grid with points $\bkf$ consists of several steps (see Fig.~\ref{fig:workflow}):
(i) Compute the $e$-d matrix elements in the Bloch representation on a coarse BZ grid with points $\bkc$ using Eq.~(\ref{eq:ed-vertex-bloch}); 
(ii) obtain the $e$-d matrix elements in the Wannier representation using Eq.~(\ref{eq:ed-b2w}); 
(iii) interpolate the Hamiltonian using Eq.~(\ref{eq:hr2hk}) and diagonalize it to obtain the fine-grid unitary matrices $U_{\bkf}$; 
(iv) interpolate the $e$-d matrix elements to any desired pair of fine-grid points $\bkpf$ and $\bkf$ using the matrix elements in the Wannier representation and the fine-grid unitary matrices [see Eq.~(\ref{eq:ed-w2b})].
%

%
%%%%%%%%%%%%%%%%%%%%%%%%%%%%%%%%%%%%%%%%%%
%%% Figure 2: M along high symmetry lines  
%%%%%%%%%%%%%%%%%%%%%%%%%%%%%%%%%%%%%%%%%%
\begin{figure}[!t]
\centering %0.85
\includegraphics[width=\linewidth]{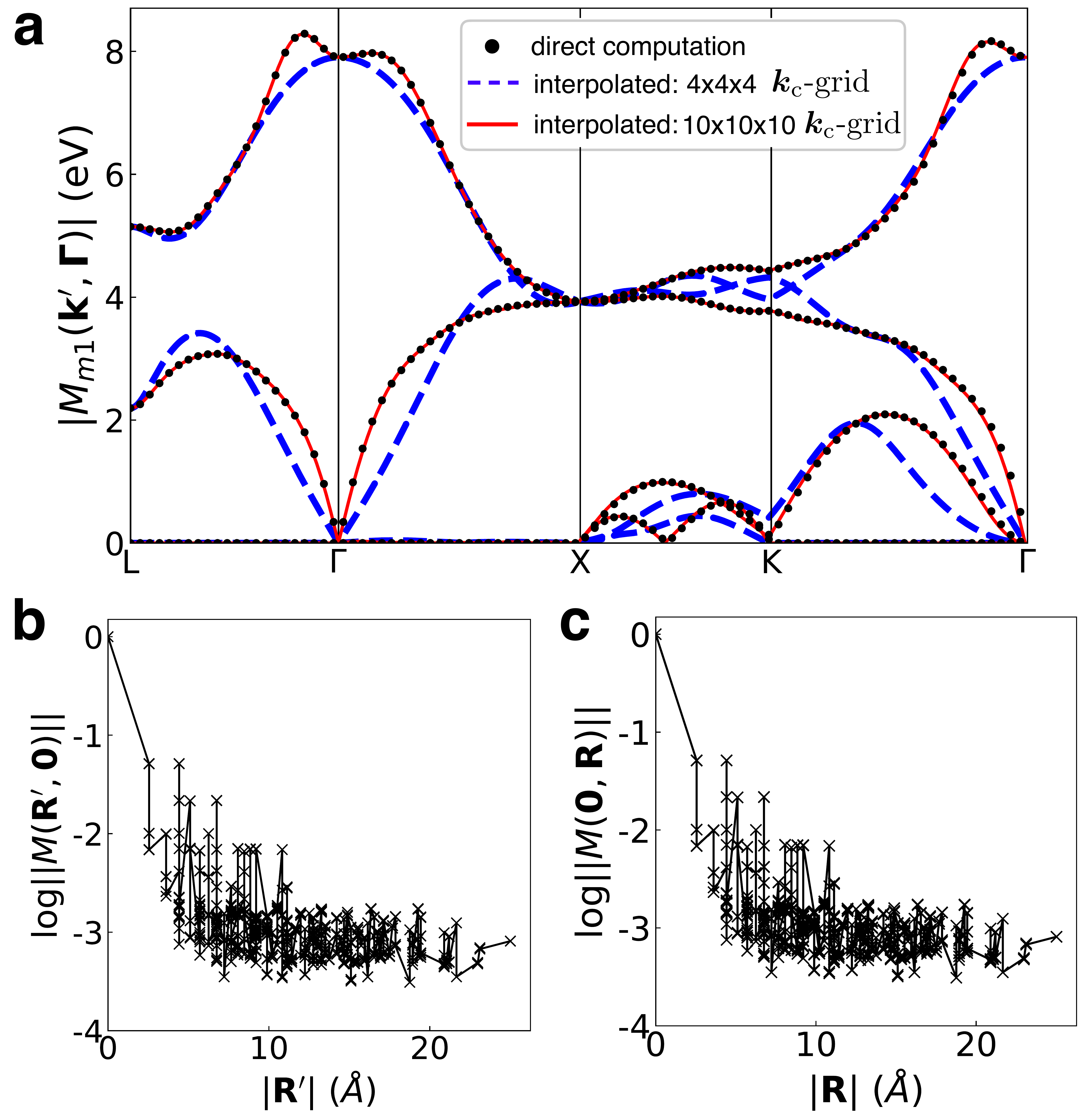}
\caption{
\textbf{a} Absolute value of the $e$-d matrix elements, computed along high-symmetry BZ lines. The initial state is set to the lowest valence band at $\mathbf{\Gamma}$, while the final states span all four valence bands and possess crystal momenta chosen along the high-symmetry lines shown in figure.
Panels \textbf{b} and \textbf{c} show the spatial decay of the $e$-d matrix elements in the Wannier basis, $||M(\bRp,\bR)||$, which are plotted in \textbf{b} as a function of $|\bRp|$ for $\bR=\mathbf{0}$ and in \textbf{c} as a function of $|\bR|$ for $\bRp=\mathbf{0}$. The highest value of $||M(\bRp,\bR)||$ is normalized to 1 in both cases, and the plots use a logarithmic scale.
}\label{fig:M_highsymlines}
\end{figure}
%

%
% INTERPOLATED E-D MATRIX ELEMENTS 
%
We validate our WF-based interpolation method using vacancy defects in silicon as an example (see Methods).
Figure~\ref{fig:M_highsymlines}a compares the $e$-d matrix elements calculated directly using Eq.~(\ref{eq:ed-vertex-bloch}) with the same matrix elements obtained by interpolation starting from two different coarse BZ grids with respectively $4^{3}$ and $10^{3}$ points $\bkc$ (here and below, we denote an $N\,\times\,N\,\times\,N$ uniform grid as $N^{3}$).
The interpolated results can qualitatively reproduce the direct computation for both coarse grids, but the results from the $10^{3}$ coarse grid achieve a superior quantitative accuracy 
as the interpolated matrix elements agree with the directly computed ones within $1\%$ over the entire BZ.
This trend implies that the $e$-d perturbation potential decays to a negligible value over more than 4 but less than 10 lattice constants.\\
\indent
%
%
% SPATIAL DECAY OF E-D MATRIX ELEMENTS IN WANNIER BASIS
%
The spatial decay of the matrix elements in the WF basis is essential for the accuracy of our approach. 
To analyze the spatial behavior of the matrix elements in the WF basis, we define for each pair of lattice vectors $\bRp$ and $\bR$ the maximum absolute value of the $e$-d matrix elements as $||M(\bRp,\bR)||=\text{max}_{ij}|M_{ij}(\bRp,\bR)|$.
Figures~\ref{fig:M_highsymlines}b and~\ref{fig:M_highsymlines}c show the spatial behavior of $||M(\bRp,\bR)||$ for a vacancy defect in silicon as a function of $|\bR'|$ while keeping $\bR=\mathbf{0}$ and as a function of $|\bR|$ while keeping $\bRp=\mathbf{0}$, respectively.
We find that the matrix elements in the WF basis decay exponentially over a few unit cells, thus confirming that the WFs are a suitable basis set for interpolating the $e$-d matrix elements.\\

\noindent
\textbf{Relaxation times and defect-limited transport}\vspace{2pt}\\
%
% WHY RTS AND MOBILITY ARE IMPORTANT
%
The $e$-d RTs and the defect-limited carrier mobility are key to characterizing carrier dynamics at low temperatures, and also near room temperature in highly doped or disordered materials. 
%
% e-d relation time formula
We compute the $e$-d RTs, $\tau_{n\bk}$, associated with elastic carrier-defect scattering using lowest-order perturbation theory~\cite{Lu2019}:
%
% EQ: RELAXATION TIME
\begin{equation}\label{eq:edtau}
%\begin{split}
\tau_{n\bk}^{-1}=\frac{2\pi}{\hbar}\frac{n_{\text{at}}C_{\rm d}}{N_{\bkp}}\sum_{m\bkp}\left|M_{mn}(\bkp,\bk)\right|^{2}\delta(\varepsilon_{m\bkp}-\varepsilon_{n\bk}),
\end{equation}
where $\hbar$ is the reduced Planck constant, $n_{\text{at}}$ the number of atoms in a primitive cell, $C_{\rm d}$ the (dimensionless) defect concentration, $N_{\bkp}$ the number of $\bk$-points used in the summation, 
and $\varepsilon_{n\bk}$ the unperturbed energy of the Bloch state $\ket{n\bk}$ in the primitive cell. 
The delta function is implemented as a normalized Gaussian with a small broadening $\eta$, $\delta_\eta (x)= e^{-x^2/2\eta^2}/\sqrt{2\pi}\eta$.
Note that the $e$-d RTs are proportional to the defect concentration since our approach assumes that the scattering events are independent and uncorrelated~\cite{Lu2019}.
For defect-limited carrier transport, we first compute the conductivity tensor $\sigma(T)$ at temperature $T$~\cite{ZhouBernardi2016} using
%
% EQ: MOBILITY
\begin{equation}\label{eq:Boltz}
\sigma_{\alpha\beta} (T) =e^{2}\int_{-\infty}^{+\infty}\!\!dE\, \left[\, -\partial f(T,E)/\partial E \,\right] \times \Sigma_{\alpha\beta}\left(E\right),
\end{equation}
where $e$ is the electron charge and $E$ the electron energy, $f(T,E)$ the Fermi-Dirac distribution, and $\Sigma\left(E\right)$ the transport distribution function (TDF) at energy $E$, defined as
\begin{equation}\label{eq:TDF}
\Sigma_{\alpha\beta}\left(E\right)=\frac{2}{\Omega_{\rm uc}}\sum_{n\bk}\tau_{n\bk}\mathbf{v}_{n\bk}^{\alpha}\mathbf{v}_{n\bk}^{\beta}\delta\left(E-\varepsilon_{n\bk}\right),
\end{equation}
where $\alpha$ and $\beta$ are Cartesian directions, and $\Omega_{\rm uc}$ the volume of the primitive cell. 
The TDF is computed with a tetrahedron integration method~\cite{ZhouBernardi2016}, using our calculated $e$-d RTs and Wannier-interpolated band velocities $\mathbf{v}_{n\bk}$~\cite{Yates2007,Mostofi2014}.
The mobility is obtained as $\mu=\sigma/n_{c}e$, where $n_c$ is the carrier concentration, while the resistivity is obtained by inverting the conductivity tensor.

%
%%%%%%%%%%%%%%%%%%%%%%%%%%%%%%%%%%%%%%%%%%
%%% FIG. 3: RELAXATION TIME AND MOBILITY  
%%%%%%%%%%%%%%%%%%%%%%%%%%%%%%%%%%%%%%%%%%
%\begin{figure}[!b]
\begin{figure}[!t]
\centering %0.85
\includegraphics[width=0.95\linewidth]{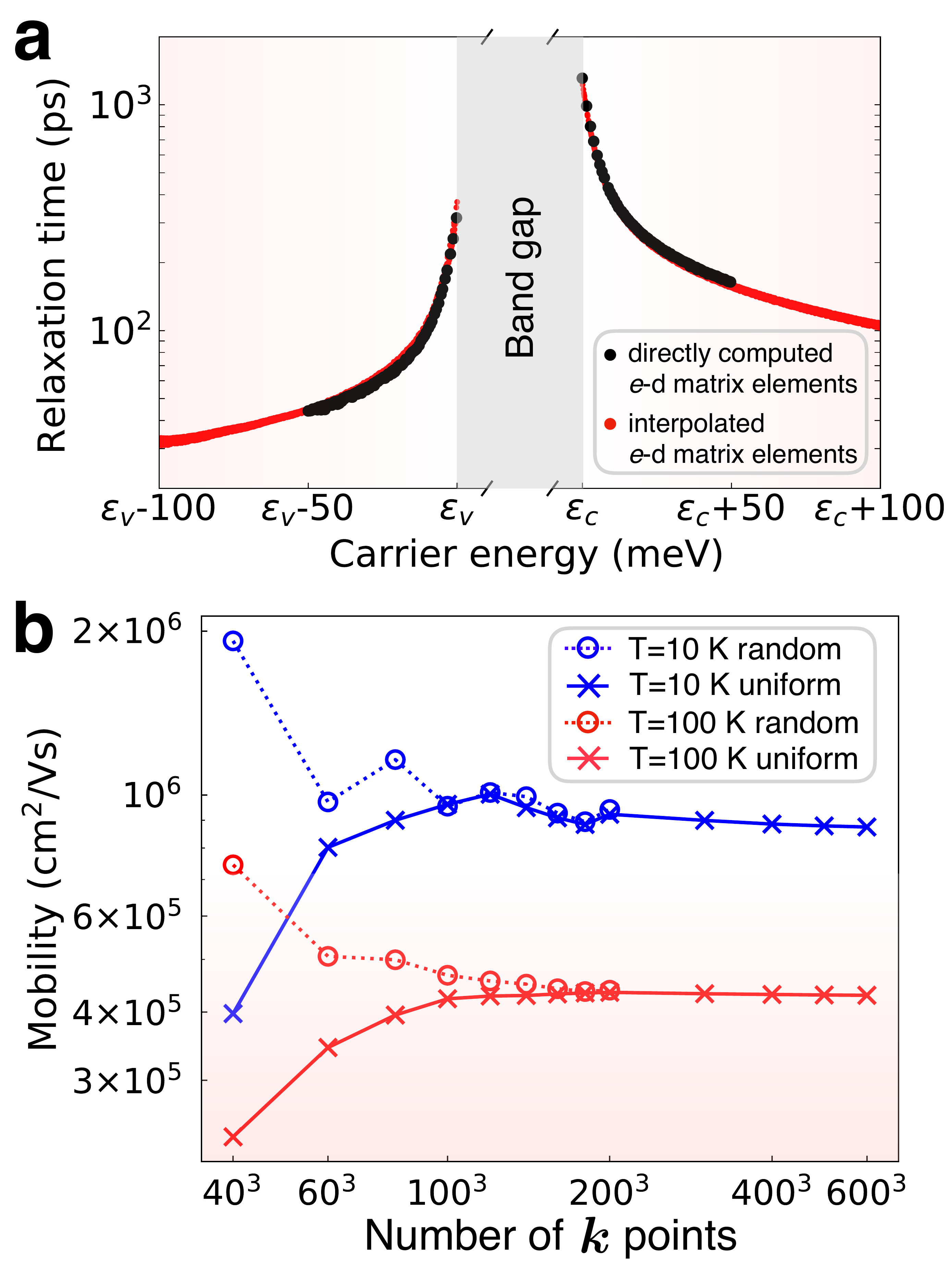}
\caption{
Carrier relaxation times and the hole mobility for vacancy defects in silicon. 
\textbf{a} Electron-defect RTs obtained from directly computed and interpolated $e$-d matrix elements. Here, $\varepsilon_{\rm c}$ is the conduction band minimum and $\varepsilon_{\rm v}$ the valence band maximum.
\textbf{b} Convergence of the hole mobility with respect to the size of the fine BZ grid used for interpolation, shown for both uniform and random grids. 
A reference vacancy concentration of $1$ ppm is used in all calculations.
}\label{fig:ed_tau_mu}
\end{figure}
%

%
% AB INITIO VERSUS WANNIER FUNCTION INTERPOLATED RELAXATION TIMES
%
We first study the $e$-d RTs and hole carrier mobility in silicon with vacancy defects (see Methods). 
We use interpolated $e$-d matrix elements, and focus on the accuracy and convergence of our interpolation method. 
The results given here assume a defect concentration of 1 vacancy in $10^{6}$ silicon atoms ($1$ ppm concentration), but results for different defect concentrations 
can be obtained by rescaling these reference RTs to any desired concentration [using Eq.~(\ref{eq:edtau})]. 
%
% relaxation times of direct and interpolation methods
Figure~\ref{fig:ed_tau_mu}a compares the RTs obtained from directly computed and interpolated $e$-d matrix elements.
The two sets of RTs are in close agreement with each other for both electrons and holes, confirming the accuracy of our interpolation method. 
%
% mobility convergence with respect to number of k points
Figure~\ref{fig:ed_tau_mu}b shows the convergence of the defect-limited hole mobility with respect to the size of the fine BZ grids, for BZ grids ranging from $40^{3}$ to $600^{3}$ points; the convergence is studied at two temperatures ($10$ and $100$ K) and for two types of grids, random and uniform.
At $10$ K, the mobilities are fully converged for fine grids with $200^{3}$ points, for both random and uniform grids. 
We observe a similar trend at $100$ K. 
The converged values of the mobility are consistent with our previous calculations using directly computed (rather than interpolated) matrix elements~\cite{Lu2019}.
%

%
% COMPARE TIME COST OF DIRECT COMPUTATION AND INTERPOLATION METHODS
%
The interpolation method allows us to use extremely dense BZ grids with up to $600^{3}$ points due to its superior computational efficiency.
Let us briefly analyze the overall speed-up of the interpolation method for a carrier mobility calculation.  
To converge the mobility at low temperature, one needs to consider only fine BZ grid points in a small energy window, roughly within $100$ meV of the band edges in semiconductors~\cite{ZhouBernardi2016, Lu2019} or of the Fermi energy in metals; these are the only states contributing to the conductivity in Eq.~(\ref{eq:Boltz}).
In this small energy window, the number of $\bk$-points is a small fraction $\alpha$ of the total number of points $N_{\bkf}$ in the entire fine BZ grid.
In the direct computation, one computes the $e$-d matrix elements $M(\bkpf,\bkf)$ between all crystal momentum pairs, and thus a number of matrix elements of order $(\alpha N_{\bkf})^{2}$. 
In the interpolation approach, the most time-consuming step is directly computing the $N_{\bkc}^{2}$ $e$-d matrix elements on the coarse grid [step (i) in Fig.~\ref{fig:workflow}], whereas interpolating the matrix elements per se is orders of magnitude less computationally expensive~\footnote{In our machine, the average CPU time to directly compute one matrix element is $\sim$ 0.2 s for silicon (with an electronic kinetic energy cutoff of 40 Ry), while the same calculation done with our interpolation method requires only $\sim$ 80 $\mu$s. The speed-up of the interpolation scheme for computing matrix elements is thus around three orders of magnitude.}. 
As a result, the overall speed-up of the interplation approach over the direct computation is $\sim (\alpha N_{\bkf})^2/N_{\bkc}^{2}$.
The typical value of $N_{\bkf}$ is around $10^{3}$$-$$10^{4} N_{\bkc}$.
For our silicon calculations, the value of $\alpha$ is of order $10^{-2}$, so the interpolation approach speeds up the mobility calculation by at least 2$-$4 orders of magnitude.
%

%
%%%%%%%%%%%%%%%%%%%%%%%%%%%%%%%%%%%%%%%%%%
%%% FIG. 4: FERMI SURFACE ADN RESISTIVITY  
%%%%%%%%%%%%%%%%%%%%%%%%%%%%%%%%%%%%%%%%%%
%\begin{figure}[!b]
\begin{figure}[!b]
\centering %0.85
\includegraphics[width=0.9\linewidth]{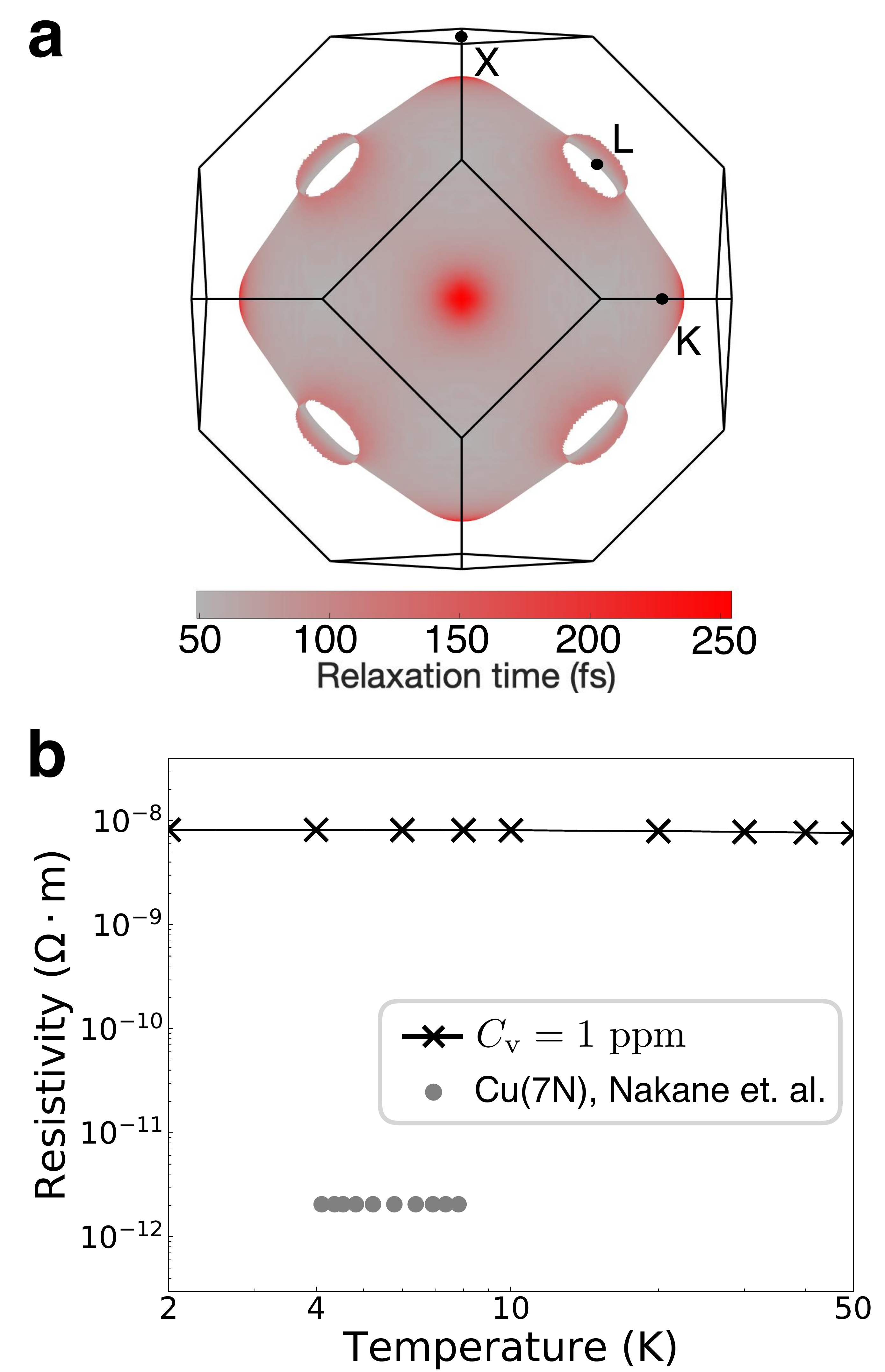}
\caption{
Relaxation times and the defect-limited resistivity for vacancy defects in copper. 
\textbf{a} RTs mapped on the Fermi surface, obtained using interpolated $e$-d matrix elements for a reference vacancy concentration of $1$ ppm.
\textbf{b} Defect-limited resistivity as a function of temperature for an assumed reference vacancy concentration of $1$ ppm, compared with experimental data from Ref.~\cite{nakane1992measuring}.
}\label{fig:edtau_rho_copper}
\end{figure}
%

%\noindent
%\textbf{Relaxation times and defect-limited resistivity in a metal}\\
%
% COPPER VACANCY CASE
%
% RELAXATION TIME ON THE FERMI SURFACE
%
The method for directly computing the $e$-d matrix elements we developed in Ref.~\cite{Lu2019} and the interpolation method shown here are general, and can be applied to metals, semiconductors and insulators.
As an example, we show a calculation on a metal, copper, containing vacancy defects (see Methods). 
In metals, the fine grids required to compute the $e$-d RTs near the Fermi energy and the resistivity are a major challenge for direct $e$-d calculations without interpolation.
Figure~\ref{fig:edtau_rho_copper}a shows the $e$-d RTs computed at $\bk$-points on the Fermi surface of copper, using interpolated $e$-d matrix elements obtained from a moderate size ($8\,\times\,8\times\,8$) coarse grid.
The $e$-d RTs for a reference vacancy concentration of 1 ppm are between 50$-$250 fs. 
Interestingly, these $e$-d RTs are orders of magnitude shorter than the electron RTs in silicon for the same vacancy concentration, which suggests that scattering due to vacancy defects in a metal is significantly stronger than in a semiconductor. 
The $e$-d RTs due to vacancy scattering in copper are strongly state-dependent $-$ we find values of order 50 fs on the majority of the Fermi surface, as well as values as large as 250 fs near the regions of the Fermi surface close to the X points of the BZ.
Similar state-dependent RTs due to impurity scattering in copper have been predicted using the all-electron Korringa-Kohn-Rostoker Green's function method~\cite{fedorov2008first,gradhand2010fully}.
These results show that our first-principles approach can access microscopic details of the $e$-d scattering processes.

%
%  COPPER RESISTIVITY
%
Figure~\ref{fig:edtau_rho_copper}b shows the calculated defect-limited resistivity for vacancy defects in copper, at low temperatures between 2$-$50 K (see Methods).
%
%The results are compared with the experimental low-temperature resistivity from Ref.~\cite{nakane1992measuring} for a pure copper sample denoted as Cu(7N), where 7N means 99.99999$\%$ purity.
%
The calculated resistivity is independent of temperature, in agreement with experimental results~\cite{nakane1992measuring,cho2010copper}, even though the conductivity formula we use [Eq.~(\ref{eq:Boltz})] depends on temperature via the Fermi-Dirac distribution.
The low-temperature defect-limited resistivity for a reference 1 ppm vacancy concentration is $\sim10^{-8}~\Omega\cdot$m.
However, the equilibrium vacancy concentration (see Methods) of copper at 50 K is negligible (of order $10^{-119}$), and the corresponding resistivity is of order $10^{-121}~\Omega\cdot$m. 
This value is negligible in comparison with the measured resistivity of $\sim10^{-12}~\Omega\cdot$m in a highly pure copper Cu(7N) sample at low temperature, where 7N means 99.99999$\%$ purity (see Fig.~\ref{fig:edtau_rho_copper}b).
We conclude that the resistivity of real copper samples at low temperature is not limited by intrinsic vacancy defects, but rather is controlled by impurities. 

Our method can predict a lower bound of the residual resistivity due to intrinsic defects in an ideally pure material. %as a function of defect concentration. %, up to concentrations at which scattering events at defects are uncorrelated.
Alternatively, if the main type of defect or impurity is known from experiment, our method can estimate the defect concentration present in the sample. 
The so-called residual resistivity ratio (RRR) between the low temperature and room temperature resistivities is used as a figure of merit for sample quality, and a large collection of data exists~\cite{hall1968survey} for RRR in metals.
Since at room temperature the resistivity is usually phonon-limited, combined with $e$-ph calculations~\cite{mustafa2016ab,ZhouBernardi2016} our approach allows one to compute RRR for a wide range of materials and defect types.
Taken together, these capabilities expand the tool box of first-principles methods for investigating carrier dynamics in complex materials.\\ 
%\\

\noindent
\textbf{DISCUSSION}\\
%
% FUTURE APPLICATIONS
%
% Charged defects
Our $e$-d interpolation method can be extended to charged defects, for which the long-range Coulomb interactions can be added in reciprocal space similar to what is done for $e$-ph interactions~\cite{sjakste2015wannier,ZhouBernardi2016}. 
%
% Spin-orbit coupling
By including spin-orbit coupling, our approach can also be extended to study spin-flip processes and $e$-d interactions in magnetic and topological materials.
%
% future applications
Future work will also attempt to include multiple $e$-d scattering events and high-order $e$-d interactions, such as those leading to localization effects in disordered systems.
Other applications include investigating carrier scattering due to extended defects such as dislocations and grain boundaries, a topic of prime relevance to materials science.
In summary, the applications of the method discussed in this work are broad, and so are its possible future extensions.

%
% CONCLUSION
%
In conclusion, we developed a WF-based interpolation approach to efficiently compute $e$-d interactions and the associated matrix elements on fine BZ grids. 
We have shown that the interpolation method is accurate and that it can effectively compute demanding BZ integrals requiring up to $10^{8}$-$10^{9}$ $\bk$-points.
The ability to efficiently interpolate $e$-d matrix elements starting from moderate BZ coarse grids is a stepping stone toward perturbative calculations of defect-limited charge and spin transport 
and to investigate quantum transport regimes governed by $e$-d interactions.\\

%
% COMPUTATIONAL DETAILS
%
\noindent
\textbf{METHODS}\vspace{2pt}\\
\textbf{DFT calculations}\vspace{2pt}\\
The ground state of a primitive cell and of supercells with size $N\times N \times N$ (where $N$ is the number of primitive cells along each lattice vector) are computed using DFT within the local density approximation.
We use a plane-wave basis set and norm-conserving pseudopotentials~\cite{van2018pseudodojo} with the {\sc Quantum Espresso} code~\cite{Giannozzi2009}. 
The total energy is converged to within 10 meV/atom in all structures.
In the defect-containing supercells, the atomic forces are relaxed to within 25 meV/{\AA} to account for structural changes induced by the defect.
For silicon, we use an experimental lattice constant of 5.43 \r{A} and a plane-wave kinetic energy cutoff of 40 Ry.
For copper, we use an experimental lattice constant of 3.61 \r{A} and a plane-wave kinetic energy cutoff of 90 Ry.
We use a $12\,\times\,12 \,\times\, 12$ $\bk$-point grid~\cite{Monkhorst1976} for the primitive cells of both materials, 
and interpolate their band structures using maximally localized WFs~\cite{Marzari1997} with the {\sc Wannier90} code~\cite{Yates2007,Mostofi2014}.\\

\noindent
\textbf{Electron-defect matrix elements, relaxation times, and resistivity calculations}\vspace{2pt}\\
The methods to directly compute the $e$-d matrix elements and from them obtain the RTs, mobility, and conductivity (or resistivity) are described in detail in Ref.~\cite{Lu2019}.
Briefly, we compute the coarse-grid $e$-d matrix elements using the wave functions of the primitive cell, and obtain the perturbation potential due to a vacancy defect using a $6\,\times\,6\,\times\,6$ supercell. 
The atomic positions around the vacancy are relaxed up to the third nearest-neighbor shell in both silicon and copper.
In the $e$-d RT and defect-limited mobility calculations in silicon, we use only electronic states in a small ($\sim$100 meV) energy window near the band edges since these are the only states contributing to the mobility~\cite{ZhouBernardi2016}; similarly, in copper we use only states within 100 meV of the Fermi energy.
In silicon, we use a broadening value $\eta= 5$ meV to compute the delta function in Eq.~(\ref{eq:edtau}), and a uniform BZ grid with $300^{3}$ points for the RTs; for the mobility, 
we use a 1 meV broadening and $e$-d matrix elements interpolated from a $10^{3}$ coarse BZ grid. 
In copper, we compute the RTs and resistivity on a fine BZ grid with $240^{3}$ points, using a 1 meV broadening and $e$-d matrix elements interpolated from a coarse $8^{3}$ BZ grid.
%
% EQUILIBRIUM VACANCY CONCENTRATION 
The equilibrium vacancy concentration at temperature $T$ in copper is estimated using~\cite{hehenkamp1992equilibrium}
\begin{equation*}
C_{\rm v}(T) = e^{-(\Delta H_{\rm v}-T\Delta S_{\rm v})/k_{\rm B} T},
\end{equation*}
where $\Delta H_{\rm v}$ and $\Delta S_{\rm v}$ are the vacancy formation enthalpy and entropy, respectively, and $k_{\rm B}$ the Boltzmann constant.
In copper, $\Delta S_{\rm v}\!=\!3.0\ k_{\rm B}$ and $\Delta H_{\rm v}\!=\!1.19$ eV~\cite{hehenkamp1992equilibrium}.\\

\noindent
\textbf{DATA AVAILABILITY}\\
The data that support the findings of this study are available from the corresponding author upon reasonable request. \\

\noindent
\textbf{CODE AVAILABILITY}\\
The code developed in this work will be released in the future at \url{http://perturbo.caltech.edu/}. \\

\noindent
\textbf{ACKNOWLEDGEMENTS}\\
This work was supported by the Air Force Office of Scientific Research through the Young Investigator Program, grant FA9550-18-1-0280.
J.-J. Z. was supported by the National Science Foundation under grant No. ACI-1642443, which provided for code development. 
J. P. acknowledges support by the Korea Foundation for Advanced Studies.\\

\noindent
\textbf{AUTHOR CONTRIBUTIONS}\\
I-T. L. derived the equations, implemented the code and carried out the numerical calculations. 
J. P. and J.-J. Z. contributed to developing the code. 
M. B. supervised the research.
M. B. and I-T. L. analyzed the results and wrote the manuscript.
All authors edited the manuscript.\\

%\section*{Additional information}
\noindent
\textbf{ADDITIONAL INFORMATION}\\
%\noindent
%\textbf{Supplementary Information} accompanies the paper on the \textit{npj Computational Materials} website (URL).\\
\noindent
\textbf{Competing interests:} The authors declare no competing interests.
%\noindent
%\textbf{Publisher's note:} Springer Nature remains neutral with regard to jurisdictional claims in published maps and institutional affiliations. 

%
% REFERENCE
%
\bibliography{wannier-interpolation}

\end{document}